\pdfoutput=1
\documentclass[aps,prl,groupedaddress,showpacs,showkeys,twocolumn,amsmath,
amsfonts,amssymb,superscriptaddress]{revtex4-1}
\usepackage{graphicx}
\usepackage[utf8x]{inputenc}
\usepackage{hyperref}
\usepackage{xcolor}
\usepackage{qcircuit}
\usepackage{physics}
\usepackage{amsbsy}

\hypersetup{
    colorlinks,
    linkcolor={red!50!black},
    citecolor={blue!60!black},
    urlcolor={blue!50!black}
}

\begin{document}


\title{Entanglement distance for an arbitrary state of $M$ qubits}


\author{Denise Cocchiarella}
\affiliation{DSFTA, University of Siena, Via Roma 56, 53100 Siena, Italy}
\author{Stefano Scali}
\affiliation{DSFTA, University of Siena, Via Roma 56, 53100 Siena, Italy}
\affiliation{Department of Physics, University of Cambridge, Cambridge CB3 0HE, United Kingdom}
\author{Salvatore Ribisi}
\affiliation{Centre de Physique Th\'eorique, Aix-Marseille University,
Campus de Luminy, Case 907,
13288 Marseille Cedex 09, France}
\author{Bianca Nardi}
\affiliation{DSFTA, University of Siena, Via Roma 56, 53100 Siena, Italy}
\author{Ghofrane Bel Hadj Aissa}
\affiliation{DSFTA, University of Siena, Via Roma 56, 53100 Siena, Italy}
\affiliation{Centre de Physique Th\'eorique, Aix-Marseille University,
Campus de Luminy, Case 907,
13288 Marseille Cedex 09, France}
\author{Roberto Franzosi}
\email[]{roberto.franzosi@ino.it}
\affiliation{QSTAR \& CNR - Istituto Nazionale di Ottica, Largo Enrico Fermi 2, I-50125 Firenze, Italy}


\date{\today}

\begin{abstract}
We propose a measure of entanglement that can be computed for any pure state of an $M$-qubit system. The entanglement measure has the form of a distance that we derive from an adapted application of the Fubini-Study metric. This measure is invariant under local unitary transformations and defined as trace of a suitable metric that we derive, the entanglement metric $\tilde{g}$. Furthermore, the analysis of the eigenvalues of $\tilde{g}$ gives information about the robustness of entanglement.

\end{abstract}


\maketitle



\section{Introduction}

Entanglement is an essential resource for progressing in the field of quantum-based technologies. Quantum information has confirmed its importance in quantum cryptography and computation, in teleportation, in the frequency standard improvement problem and metrology based on quantum phase estimation \cite{GUHNE20091}. The rapid experimental progress on quantum control is driving the interest in entanglement theory. Nevertheless, despite its key role, entanglement remains elusive and the problem of its characterisation and quantification is still open \citep{PhysRevA.95.062116,PhysRevA.67.022320}. We propose a measure of entanglement that can be computed on any pure state of M-qubit systems. The measure is derived from a tailored form of the Fubini-Study metric that we verify to correspond to the quantum Fisher information but which allows for a deeper understanding thanks to its eigenvalues' analysis. The measure that we propose: i) is invariant under local unitary transformations; ii) has the structure of a distance such that the higher is the entanglement of a given state the greater is its minimum distance from infinitesimally close states; iii) distinguishes between the case $M=2$ and the case $M>2$ which seems to be consistent with the fact that most of the propositions which are necessary and sufficient in the case $M=2$ lose the sufficient condition in the $M>2$ case.

\section{Distance entanglement}
The Hilbert space of an $M$-qubit system carries the
Fubini-Study metric \cite{gibbons}
\begin{equation}
\langle d \psi | d \psi \rangle - \dfrac{1}{4}
|\langle \psi | d \psi \rangle - \langle d \psi | \psi \rangle|^2 \, ,
\label{F-S-metric}
\end{equation}
where $|\psi \rangle$ is a generic normalised state and $|d\psi \rangle$
is an infinitesimal variation of such state.
\emph{The present study is aimed to endow the Hilbert space with
a Fubini Study-like metric that has the desirable property
of making it an attractive definition for entanglement
measure.}
For this reason, such distance should not be affected by
local operations on single qubits.
As a matter of fact, the action of $M$ arbitrary $SU(2)$ local unitary 
operators $U^j$ ($j=0,\ldots,M-1$) on a given state $|s\rangle$,
generates a class of states
\begin{equation}
|U,s\rangle = \prod^{M-1}_{j=0} U^j |s\rangle
\label{Us}
\end{equation}
that share the same degree of entanglement.
For each $j$, $U^j$ operates on the $j$th qubit.
We define an infinitesimal variation of state \eqref{Us} as
\begin{equation}
|dU,s\rangle = \sum^{M-1}_{j=0} d\tilde{U}^j |U,s\rangle \, ,
\label{dUs}
\end{equation}
where
\begin{equation}
d\tilde{U}^j = -i 
({\bf n}^j \cdot \boldsymbol{\sigma}^j)
d \xi^j/2 
\label{dU}
\end{equation}
rotates the $j$th qubit by an infinitesimal angle $d\xi^j$  around
the unitary vector ${\bf n}^j $.
We denote by $\sigma^j_1$, $\sigma^j_2$ and $\sigma^j_3$
the Pauli matrices operating on the $j$-th qubit ($j=0,\ldots,M-1$)
where the index $j$ numerates the spins from right to left.
From Eq. \eqref{F-S-metric}, with this choice, we get the following
expression for the Fubini-Study metric $g$
\begin{align}
g_{\mu \nu} ({\bf v}^\nu) d\xi^\mu d\xi^\nu & =
\dfrac{1}{4}
\left(
\langle s | ({\bf v}^\mu \cdot \boldsymbol{\sigma}^\mu)
({\bf v}^\nu \cdot \boldsymbol{\sigma}^\nu)|s\rangle +
\right. \nonumber
\\ &
\left.
- 
\langle s | ({\bf v}^\mu \cdot \boldsymbol{\sigma}^\mu)
|s\rangle
\langle s |
({\bf v}^\nu \cdot \boldsymbol{\sigma}^\nu)|s\rangle
\right) d\xi^\mu d \xi^\nu \, .
\label{gmunu}
\end{align}
The unitary vectors ${\bf v}^j$ in the latter equation are
derived by a rotation of the original ones according to
\begin{equation}
{\bf v}^\nu \cdot \boldsymbol{\sigma}^\nu =
U^{\nu \dagger} {\bf n}^\nu \cdot \boldsymbol{\sigma}^\nu
U^{\nu} \, ,
\end{equation}
where there is no summation on the index $\nu$.
\emph{The proposed entanglement measure of the state $|s\rangle$
is }
\begin{equation}
E(|s\rangle) =\inf_{{\cal M}\ \{{\bf v}^\nu\}_\nu} \tr (g) \, ,
\label{emeasure}
\end{equation}
\emph{where $\tr$ is the trace operator and where the $\inf$ is taken
``in measure'' over all the possible orientations of the unitary
vectors $ {\bf v}^\nu$.}
With the term ``in measure'', we mean that possible pathologies,
similar to the one of the Dirichlet function, are eliminated.
\emph{The $\inf$ operation, makes the measure \eqref{emeasure}
independent from the operators $U_j$ hence, its numerical
value is associated to the class of states generated by local unitary transformations and not to the specific element
chosen inside the class. This is a necessary condition
for a good entanglement measure definition.}
The unitary vectors $\tilde{\bf v}^\nu$ corresponding
to the $\inf$ of $\tr(g)$, identify a metric
\begin{equation}
\tilde{g} = g(\tilde{\bf v}^\nu)
\label{EM}
\end{equation}
that we name entanglement metric (EM).
The off-diagonal elements of $\tilde{g}$ provide
the quantum correlations between qubits.
In addition, states that differ one another for local unitary
transformations, have the same form of $\tilde{g}$.
In this way, the expression of EM identifies the classes of
equivalence.
Remarkably, the analysis of the eigenvalues and
eigenvectors of $\tilde{g}$ allows one to
check the existence of states with
super-Heisenberg sensitivity, i.e. beyond
Heisenberg limit.

\section{Examples}
In order to verify the efficacy of the proposed entanglement
measure, we
have first considered two families of one-parameter states
depending on a real parameter.
The degree of entanglement of each state depends on this
parameter and the configuration corresponding to the maximally entangled states for each of the families is known.
The first family of states we consider has been introduced by 
Briegel and Raussendorf in Ref. \cite{briegel_PRL86_910}. For this
reason, we will name the elements in this family
Briegel-Raussendorf states (BRS).
The second family of states is related to
the  Greenberger-Horne-Zeilinger states \cite{ghz}.
We will name the elements of such family Greenberger-Horne-Zeilinger--like
states (GHZLS).
\emph{It is worth emphasizing that in Ref. \cite{briegel_PRL86_910} it has
been shown that the maximally entangled states of these two families are
not equivalent if $M \geq 4$, whereas they are equivalent if $M\leq 3$.
This fact offers us a further test for our approach to entanglement estimation.
In fact, we have found that i) the entanglement measure \eqref{emeasure} provides the
same value for the maximally entangled states of both the families;
ii) in the case $M\leq 3$ the entanglement metric \eqref{EM} has the same
form for the maximally entangled states of the two families,
whereas, if $M \geq 4$, the EMs of the maximally entangled states
of the two families are not equivalent.}

The last case we have considered is a family of three-qubit states depending
on two real parameters. With a suitable choice of these parameters, the state
can be fully separable or bi-separable, whereas in the generic case it is a
genuine tripartite entangled state. We will show that the proposed
entanglement measure provides an accurate description of all these
cases.

\subsection{Briegel Raussendorf states}
We denote with $\Pi^j_0=(\mathbb{I}+\sigma^j_3)/2$
and $\Pi^j_1=(\mathbb{I}-\sigma^j_3)/2$ the projector operators
onto the eigenstates of $\sigma^j_3$, $|0\rangle_j$ (with
eigenvalue $+1$)
and $|1\rangle_j$ (with eigenvalue $-1$),
respectively.
Each $M$ qubit state of the BRS class is
derived by applying to the fully separable state
\begin{equation}
|r, 0 \rangle = \bigotimes^{M-1}_{j=0}
\dfrac{1}{\sqrt{2}}(|0\rangle_j + |1\rangle_j)
  \, ,
\label{r0i}
\end{equation}
the non local unitary operator 
\begin{equation}
U_0(\phi) = \exp (-i \phi H_0) = \prod\limits^{M-1}_{j=1}
\left(
\mathbb{I} + \alpha \Pi^j_0 \Pi^{j+1}_1 
\right) \, ,
\label{U0e}
\end{equation}
where 
$
H_0 = \sum^{M-1}_{j=1}\Pi^j_0 \Pi^{j+1}_1 
$
and 
$
\alpha = (e^{-i\phi} -1) \, .
$
The full operator \eqref{U0e} is diagonal on the states of the standard
basis
$\{
|0 \cdots  0 \rangle \, , \, \,
|0 \cdots 0 1 \rangle
, \ldots ,
|1 \cdots 1 \rangle \}
$.  In fact,
each vector of the latter basis is identified by $M$ integers
$n_0,\ldots , n_{M-1} =0,1$ as
$
\ket{\{n\}} = |n_{M-1} \ n_{M-2} \quad n_0 \rangle \, 
$
and we can enumerate such vectors according to the binary integers representation
$ |k\rangle = \ket{\{n^k\}}$, with $k = \sum^{M-1}_{j=0} n^k_j 2^{j} $,
where $n^k_\nu$ is the $\nu$-th digit
of the number $k$ in binary representation and $k=0,\ldots,2^M-1$.
Then, the eigenvalue $\lambda_k$ of operator \eqref{U0e}, corresponding to a given
eigenstate $|k\rangle$ of this basis, results 
\begin{equation}
\lambda_k  = \sum^{n(k)}_{j=0} \binom{n(k)}{j} \alpha^j \, ,
\end{equation}
where $n(k)$ is the number ordered couples $01$ inside the sequence of the base
vector $|k\rangle$.
For the initial state \eqref{r0i} we consistently get 
\begin{equation}
|r, 0 \rangle_M =  2^{-M/2} \sum^{2^M-1}_{k=0} |k\rangle \, ,
\label{r0}
\end{equation}
and, under the action of $U_0(\phi)$ one
obtains
\begin{equation}
\begin{aligned}
|r, \phi \rangle_M = 
2^{-M/2} \sum^{2^M-1}_{k=0} 
\sum^{n(k)}_{j=0} \binom{n(k)}{j} \alpha^j
|k\rangle
 \, .
\label{state-phi}
\end{aligned}
\end{equation}
%
For $\phi =2 \pi k$, with $k\in \mathbb{Z}$, this state
is separable, whereas, for all the other choices of the
value $\phi$, it is entangled.
In particular, in \cite{briegel_PRL86_910} it is argued
that the values $\phi = (2k+1)  \pi$, where $k\in \mathbb{Z}$,
give the maximally entangled states.

\subsection{Fubini-Study metric for the Briegel Raussendorf states $M=2,3$}

In the case of two-qubit BRS the trace of the Fubini-Study metric is
\begin{equation}
\tr (g) = \sum^1_{\nu=0}\left[1 - c^2\left(cv_1^\nu+\left(-1\right)^{\nu+1}sv_2^\nu\right)^2
 \right]/4 \, ,
\label{g2}
\end{equation}
where $c= \cos\left({\phi}/{2}\right)$ and $s= \sin\left({\phi}/{2}\right)$. \eqref{g2} is minimised with the choice $\tilde{\bf v}^\nu=\pm (c,(-1)^{\nu+1}s,0)$.
Consistently, the EM results in
\begin{equation}
\tilde{g} =\dfrac{1}{4}
\left(
\begin{array}{cc}
s^2 & 1\\
1 & s^2
\end{array}
\right)
\label{gtilde2}
\end{equation}
and 
\begin{equation}
E(|r, \phi \rangle_2) = \dfrac{s^2}{2} \, .
\label{Erphi2}
\end{equation}
We have already mentioned that in the case $M=2,3$,
the maximally-entangled BRS $|r, 2\pi k + \pi\rangle$, where $k\in \mathbb{Z}$, and the maximally entangled GHZLS are equivalent
because differing just for
local unitary transformations. In the following, we will show that the EM for
these states have the same forms in the case $M=2,3$ in accordance to
the results of Ref. \cite{briegel_PRL86_910}.
In the case $M=3$ and $\phi \neq (2k+1)  \pi$, with $k\in \mathbb{Z}$, the trace of $g$,
\begin{equation}
\tr (g) =\left[3 - c^2\left(c(v_1^0 + v_1^1 + v_1^2)+s(v_2^2-v_2^0)\right)^2
 \right]/4 \, ,
\label{g3}
\end{equation}
is minimised with the choices $\tilde{\bf v}^0=(c,-s,0)$, $\tilde{\bf v}^1=(1,0,0)$ and
$\tilde{\bf v}^2=(c,s,0)$.
The EM and the entanglement measure in this case results to be
\begin{equation}
\tilde{g} =\dfrac{s^2}{4}
\left(
\begin{array}{ccc}
1 & c & -2 s^2 c^2\\
c & 1 +c^2 & c \\
- 2 s^2 c^2& c & 1
\end{array}
\right)
\label{gtilde3}
\end{equation}
and
\begin{equation}
E(|r, \phi \rangle_3) = \dfrac{s^2}{4} \left( 3+ c^2 \right) \, ,
\label{Erphi2}
\end{equation}
respectively.
By direct calculation, one can verify that in the case of the maximally
entangled BRS ($M=3$), the choice ${\bf v}^0=(-1,0,0)$, ${\bf v}^1 = (0,0,1)$
and ${\bf v}^2=(1,0,0)$ makes the EM equivalent to the one of the three-qubit
Greenberger-Horne-Zeilinger state. This agrees with the results of
Ref. \cite{briegel_PRL86_910}.


\subsection{Fubini-Study metric for the Briegel Raussendorf states $M>3$}

In the general case, the trace of $g$ results
\begin{equation}
\begin{aligned}
\tr (g) &=\dfrac{1}{4}\left\{ M - \sum^{M-1}_{\nu=0} \left[
v^\nu_3
w^\nu_3 +
v^\nu_+ w^\nu_- + v^\nu_- 
 w^\nu_+
\right]^2
 \right\} \, ,
 \end{aligned}
\label{gM}
\end{equation}
where $v^\nu_\pm = v^\nu_1 \pm i v^\nu_2$,
$c_k = 2^{-M/2} \lambda_k $, and 
\begin{equation}
\begin{array}{l}
w^\nu_- = \sum^{2^M-1}_{k=0} 
\delta_{n_\nu^k,0} 
c^*_{k+2^{\nu}} c_k \, , 
\\
w^\nu_+ = \sum^{2^M-1}_{k=0} 
\delta_{n_\nu^k,1} 
c^*_{k-2^{\nu}} c_k \, , \\
w^\nu_3 = \sum^{2^M-1}_{k=0} 
(-1)^{n_\nu^k} |c_k|^2 \, .
\end{array}
\label{wM}
\end{equation}
The trace is minimised by setting  
$\tilde{v}^{\nu}_+ ={{w}^{\nu \star}_-}/{\Vert 
{\bf w}^{\nu} \Vert}$, $\tilde{v}^{\nu}_- = {{w}^{\nu \star}_+}/{\Vert 
{\bf w}^{\nu} \Vert}$ and $\tilde{v}^{\nu}_3 = \frac{{w}^{\nu}_3}{\Vert 
{\bf w}^{\nu} \Vert}$.
From the latter we get the entanglement measure for the BRS that is
\begin{equation}
E(|r, \phi \rangle_M) = \dfrac{1}{4} \left( M- 
\sum^{M-1}_{\nu=0} 
\Vert {\bf w}^\nu \Vert^2 \right) \, .
\label{ErphiM}
\end{equation}

\subsection{Greenberger-Horne-Zeilinger--like states}
Now, we consider a second class of states (GHZLS) defined according to
\begin{equation}
|GHZ,\theta \rangle_M = \cos(\theta) |0\rangle + \sin(\theta) 
e^{i \varphi}|2^M-1\rangle \, .
\label{ghz}
\end{equation}
For $\theta = k \pi/2$, where $k\in \mathbb{Z}$, these states are fully separable, whereas
$\theta = k \pi/2+\pi/4$ selects the maximally entangled 
states.
In this case, the trace for the Fubini-Study metric,
\begin{equation}
\begin{aligned}
\tr (g) &=\dfrac{1}{4}\left[ M - \cos^2(2\theta)\sum^{M-1}_{\nu=0} 
(v^{\nu }_3)^2
 \right] \, ,
 \end{aligned}
\label{g-ghzM}
\end{equation}
is minimised by the values $v^{\nu }_3 =1$. Consistently, we have
\begin{equation}
\tilde{g} = \dfrac{1}{4}\sin^2(2\theta) J_M
\end{equation}
where $J_M$ is the $M\times M$ matrix of ones. The entanglement measure
for the GHZLS results
\begin{equation}
E(|GHZ,\theta\rangle_M) = \dfrac{M}{4}  \sin^2(2\theta) \, .
\label{ErGHZ}
\end{equation}

\subsection{Three-qubit states depending on two parameters}
The last class of states we consider is
\begin{equation}
\begin{aligned}
|\varphi,\gamma, \tau \rangle_3 = &\cos(\gamma) |0\rangle
[\cos(\tau) |00\rangle +
\sin(\tau) |11\rangle ] \\
+& \sin(\gamma) |1\rangle
[\sin(\tau) |00\rangle
+ \cos(\tau) |11\rangle ] \, .
\label{3qb}
\end{aligned}
\end{equation}
These states are fully separable for $\gamma = 0,\pi/2$ and $\tau = 0,\pi/2$ whereas they are bi-separable for $\tau = \pi/4$.
In this case, the trace of the Fubini-Study metric is
\begin{equation}
\begin{aligned}
\tr (g) &=\dfrac{1}{4}\left\lbrace  3 - \cos^2(2\gamma) \cos^2(2\tau)[(v^{0}_3)^2 +
(v^{1}_3)^2 ] \right.\\
&\left.
 - 
[\sin(2\gamma) \sin(2\tau) v^{2}_1 + \cos(2\gamma)  v^{2}_3]^2
 \right\rbrace
 \end{aligned}
\label{g-3qb}
\end{equation}
and it is minimised by the values $\tilde{\bf v}^{\nu }_3 =(0,0,1)$, $\nu=0,1$ and
\begin{equation}
\begin{aligned}
\tilde{v}^{3 }_1 &=\dfrac{\sin(2\gamma) \sin(2\tau)}{\sqrt{
\sin^2(2\gamma) \sin^2(2\tau)+
\cos^2(2\gamma)}} \, ,\\
\tilde{v}^{3 }_2 &=
0 \, , \\
\tilde{v}^{3 }_3 &=\dfrac{\cos(2\gamma)}{\sqrt{
\sin^2(2\gamma) \sin^2(2\tau)+
\cos^2(2\gamma)}} \, .
\end{aligned}
\end{equation}
Consistently, the entanglement measure
for these states results to be
\begin{equation}
E(|\varphi,\gamma, \tau \rangle_3) = 
\dfrac{1}{4} [ 2 \sin^2(2\tau) +3 \sin^2(2\gamma) \cos^2(2\tau) ]\, .
\label{Er3qb}
\end{equation}

\section{Results}

\subsection{Entanglement measure}
\begin{figure}[h]
\begin{center}
{ 
\includegraphics[width=1\linewidth]{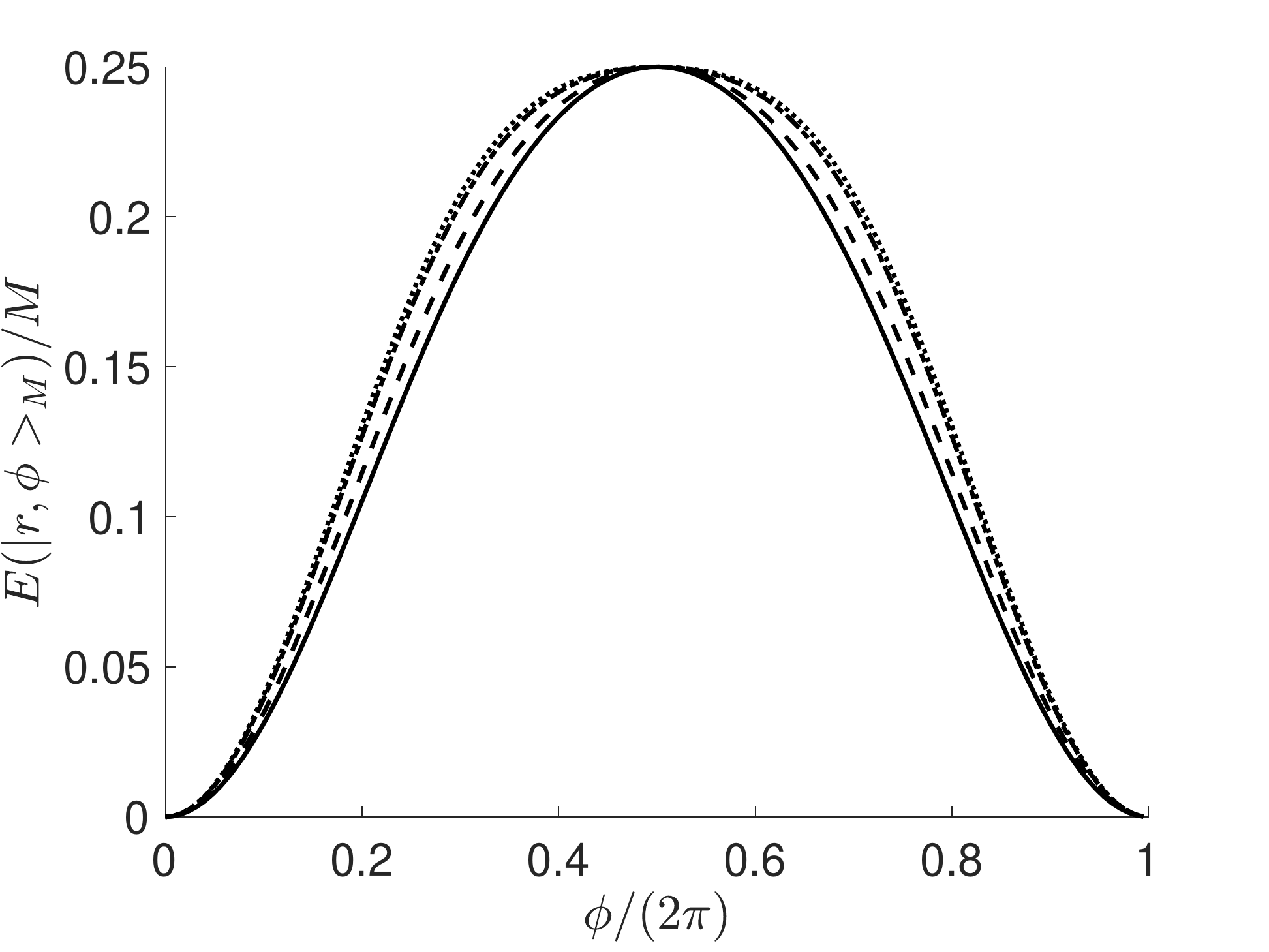}
}
\end{center}
\caption{$E(|r, \phi \rangle_M)/M$ vs $\phi/(2\pi)$  for the BRS the cases $M=3$ (continuous line),
$M=4$ (dashed line), $M=7$ (dot-dashed line) and $M=9$ (dotted line).
}
\label{Fig01}
\end{figure}

\begin{figure}[h]
\begin{center}
{ 
\includegraphics[width=1\linewidth]{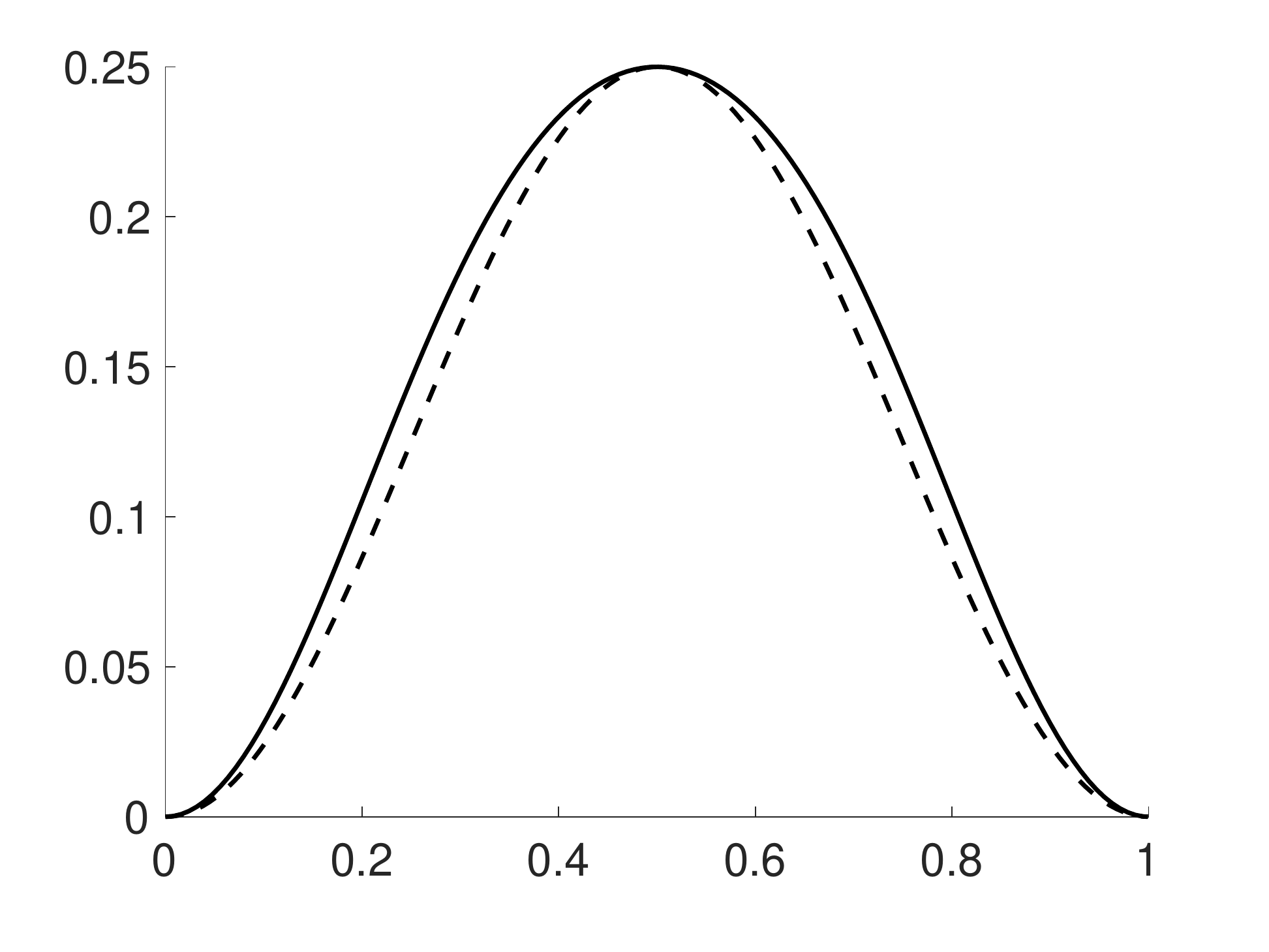}
}
\end{center}
\caption{$E(|r, \phi \rangle_M)/M$ vs $\phi/(2\pi)$ 
in continuous line and $E(|GHZ, \theta \rangle_M)/M$ vs $2 \theta/\pi$ in dashed line
for the case $M=3$.
}
\label{Fig02}
\end{figure}

In  Fig. \ref{Fig01} we plot the measure $E(|r, \phi \rangle_M)/M$ vs $\phi/(2\pi)$
according to Eq. \eqref{ErphiM}, for the  states \eqref{state-phi} in the case $M=3,4,7,9$. 
Figure \ref{Fig01} show that the proposed entanglement measure provides in all these
cases a correct estimation of the degree of entanglement for the BRS. In particular,
for the fully separable states ($\phi=0$) it gives a vanishing value, whereas for the
maximally entangled states ($\phi=\pi$) it provides the maximum possible value
for the trace, that is $E(|r, \pi \rangle_M)/M=1/4$. This implicitly indicates that on
the maximally entangled states the expectation values  for all
$\tilde{\bf v}^\nu \cdot \boldsymbol{\sigma}^\nu $ ($\nu=0,\ldots,M-1$) vanish.
The entanglement measure \eqref{emeasure} successfully passes also the second test
of the GHZLS for which it provides zero in the case of fully separable states
($\theta=0$) and the maximum value
 ($1/4$) in the case of the maximally entangled state ($\theta=\pi/2$). 
In figure \ref{Fig02} we compare the curves $E(|r, \phi \rangle_M)/M$ vs $\phi/(2\pi)$ 
in continuous line and $E(|GHZ, \theta \rangle_M)/M$ vs $2 \theta/\pi$ in dashed line
for the case $M=3$.
Also in this case, for the maximally entangled states the expectation
value for the operators $\tilde{\bf v}^\nu \cdot \boldsymbol{\sigma}^\nu $ ($\nu=0,\ldots,M-1$) is zero.

\begin{figure}[h]
\begin{center}
{ 
\includegraphics[width=1\linewidth]{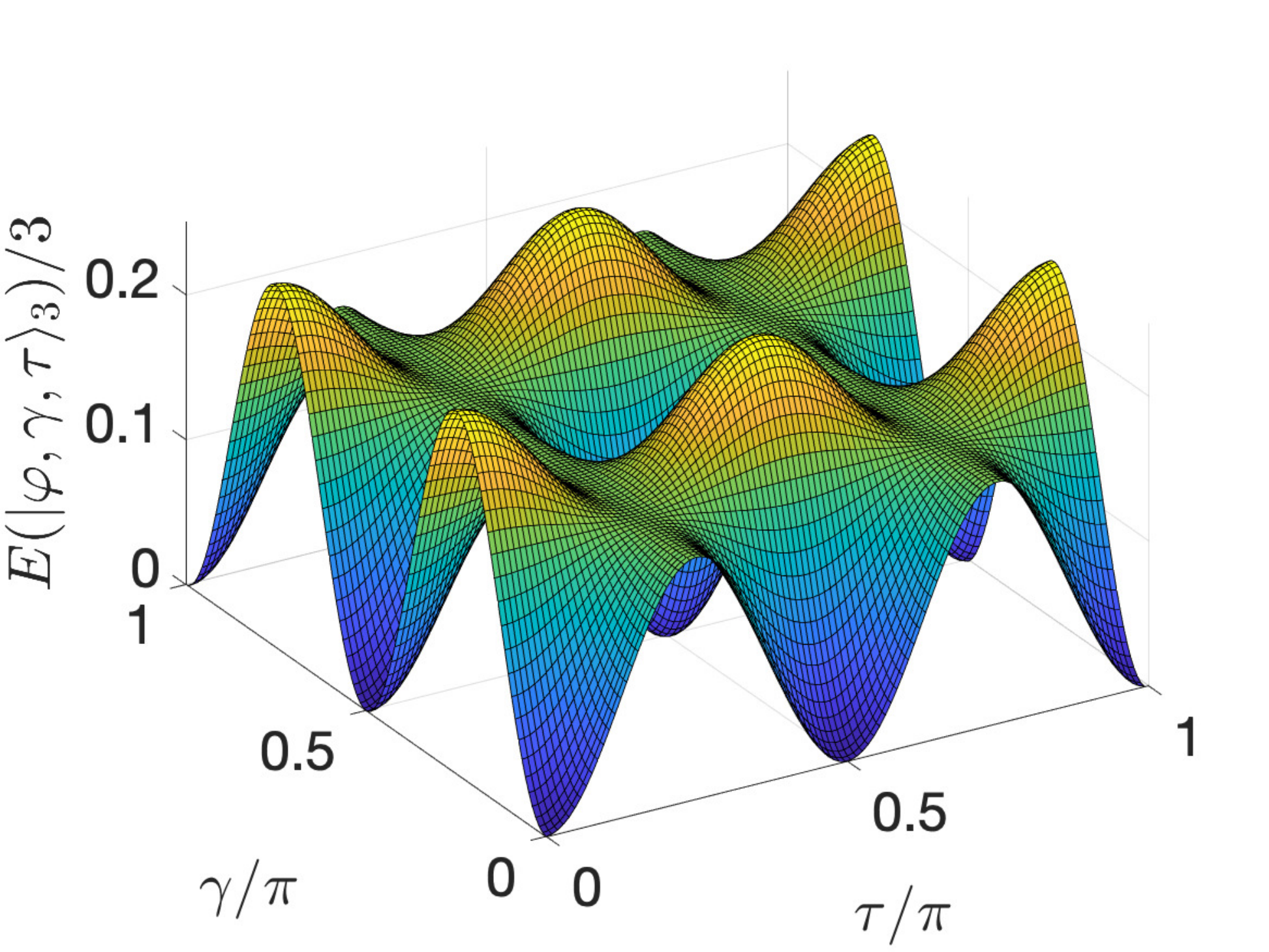}
}
\end{center}
\caption{Three dimensional plot of $E(|\varphi,\gamma,\tau \rangle_3)/3$ 
as a function of $\gamma/\pi$ and $\tau/\pi$  for the state \eqref{3qb}.
}
\label{Fig05}
\end{figure}

In  Fig. \ref{Fig05}, we report, with a 3D plot, the measure $E(|\varphi,\gamma,\tau \rangle_3)/3$ 
as a function of $\gamma/(2\pi)$ and $\tau/(2\pi)$ according to Eq. \eqref{Er3qb}, for the  states \eqref{3qb}.
The measure \eqref{emeasure} catches, in a surprisingly clear way, the entanglement
properties of this family of states. In particular, $E(|\varphi,\gamma,\tau \rangle_3)/3$
is null in the case of fully separable states ($\gamma=0,\pi/2,\pi$ and $\tau=0,\pi/2,\pi$)
and it is maximum (with value $1/4$) in the case of maximally entangled states ($\gamma = \pi/4,
3\pi/4$ and $\tau= 0,\pi/2,\pi$). In addition, the case of bi-separable states ($\tau=\pi/4$) results in $0 < E(|\varphi,\gamma,\tau \rangle_3)/3 < 1/4$.

\subsection{Eigenvalues analysis}

\begin{figure}[h]
\begin{center}
{ 
\includegraphics[width=1\linewidth]{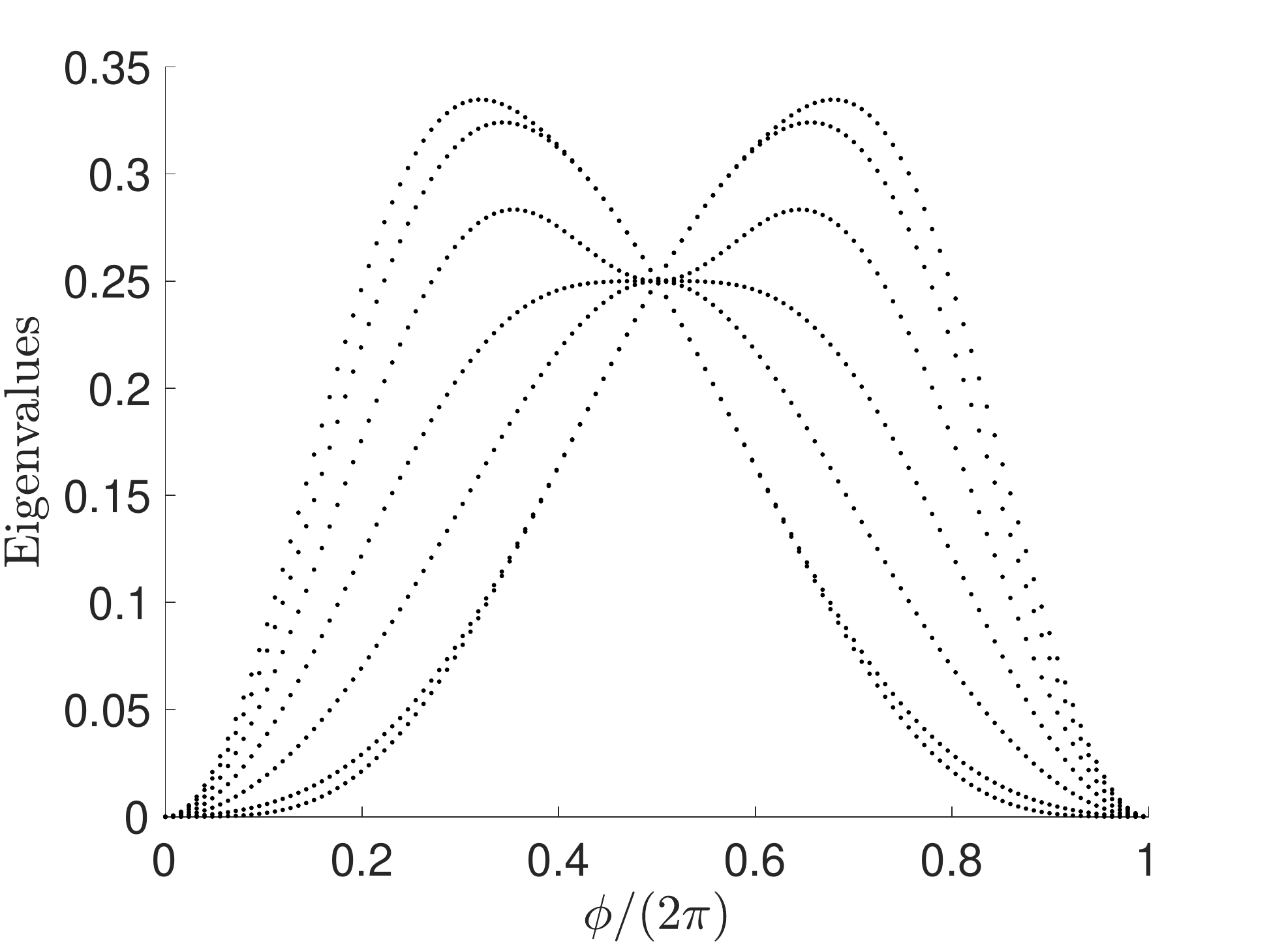}
}
\end{center}
\caption{The figure plots the $\tilde{g}$ eigenvalues for the state $|r, \phi \rangle_M$ vs
$\phi/(2\pi)$ for the case $M=7$.
}
\label{Fig03}
\end{figure}

\begin{figure}[h]
\begin{center}
{ 
\includegraphics[width=1\linewidth]{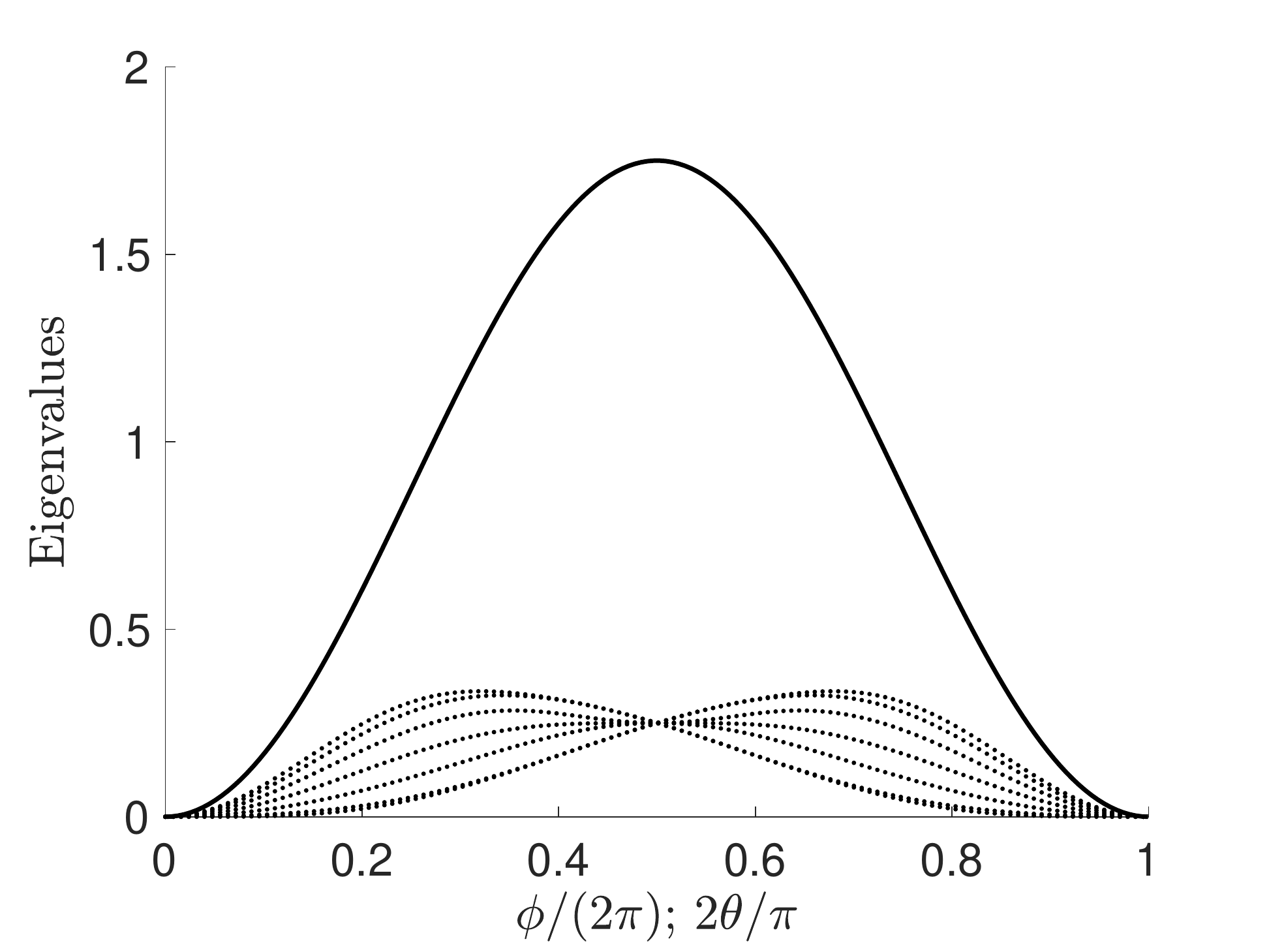}
}
\end{center}
\caption{The figure plots the $\tilde{g}$ eigenvalues for the state $|r, \phi \rangle_M)$ vs
$\phi/(2\pi)$ in dotted lines and the  unique not vanishing eigenvalue of 
$\tilde{g}$ for the state GHZLS vs $2 \theta/\pi$ in continuous line, for the case $M=7$.}
\label{Fig04}
\end{figure}

Other interesting characteristics of the entanglement measure come from the analysis of the metric's eigenvalues. In fig. \ref{Fig03}, we plot the eigenvalues of $\tilde{g}$ for the state $|r, \phi \rangle_M$ vs
$\phi/(2\pi)$ for the case $M=7$.
In the general case $\phi\neq 0, 2\pi$ the BRS $\tilde{g}$ have $M$ not null eigenvalues. 
This fact makes the class of the BRS robust, concerning
entanglement, inasmuch the minimum distance between states in a
direction randomly chosen is greater than the minimum eigenvalue. On the
contrary, the GHZLS have only one non-vanishing eigenvalue. Although the
value of the latter is greater than the eigenvalues of the BRS (see Fig.
\ref{Fig04}), the GHZLS appear weak, in the sense of entanglement, since there exist $M-1$
directions with null minimum distance between states.
In fig. \ref{Fig04}, we compare the plots of the eigenvalues of $\tilde{g}$ for $|r, \phi \rangle_M$ vs $\phi/(2\pi)$ (dotted lines), with the plot of the unique not vanishing eigenvalue of $\tilde{g}$ for GHZLS vs $2 \theta/\pi$ (continuous line), in the case $M=7$.
\vspace{10pt}

\emph{Within the scenario that we have proposed, the entanglement has the physical interpretation of an obstacle to the minimum distance between infinitesimally close states}. In fact, by defining the distance between a given state represented by the vector $|U,s\rangle$ and
an infinitesimally close state associated with the vector $|d U,s\rangle$ as 
$ds^2 =\tr(g({\bf v}))dr^{2}$
where $\sum_\mu (d\xi^{\mu })^2 = dr^2$, it results
\begin{equation}
ds^2 \geq E(|s\rangle)dr^2 \, .
\end{equation}
This shows that the minimum density distance $ds^2/dr^2$, obtained by varying the vectors ${\bf v}$, is bounded from below by the entanglement measure $E(|s\rangle)$.
For fully separable states, the minimum density distance is zero whereas for
maximally entangled states, it results $M/4$ at the very best.
It is worth emphasizing that $ds^2$ can overcome the value of $E(|s\rangle)dr^2$.

\section{Concluding remarks}

In this paper, we have introduced a new measure of entanglement for the case of an arbitrary pure state of M qubits \eqref{emeasure}. We verified the invariance under local unitary transformations identifying classes of equivalence of states, a demanded property of a good entanglement measure. Furthermore, the measure has the characteristics of a distance and assumes the intuitive physical interpretation of an obstacle to the minimum distance between infinitesimally close states. Finally, the analysis of the eigenvalues allows
one to determine if there are any states which are more sensitive to small
variations than others. For instance, Fig. \ref{Fig03} shows that, in the case of $|r,\pi/2\rangle_7$ state, a small
variation along the eigenvector's direction of the maximum eigenvalue
of $\tilde{g}$ brings a greater distance than the one
derived in the case of the maximally entangled state $|r,\pi\rangle_7$. 
This analysis is a possible useful mean in the task of determining states with super-Heisenberg sensitivity.\\

\begin{acknowledgments}
We are grateful to A. Smerzi and L. Pezz\'e for useful discussions.\\

R. F. thanks the support by the QuantERA project 
“Q-Clocks” and the European Commission.
\end{acknowledgments}


\hfill
\bibliography{references}

\end{document}